\documentclass[aps,pra,graphicx,9pt,twocolumn,superscriptaddress]{revtex4-2}
\usepackage[colorlinks,linkcolor=blue,urlcolor=blue,anchorcolor=blue,citecolor=blue]{hyperref}
\usepackage{amsmath}
\usepackage{graphicx}
\usepackage{dcolumn}
\usepackage{mathrsfs}
\usepackage{physics}
\usepackage{amssymb}
\usepackage{amsfonts,multirow}
\usepackage{bm}
\usepackage{changes}
\definechangesauthor[name={HLZhang}, color=red]{HL}
\UseRawInputEncoding
\begin{document}

\title{Scheme for measuring topological transitions in a continuous variable system}


\author{Bi-Yao Wang}
\author{Hao-Long Zhang}
\author{Shou-Bang Yang}
\author{Fan Wu}
\author{Zhen-Biao Yang*}
\email{zbyang@fzu.edu.cn}
\author{Shi-Biao Zheng*}
\email{t96034@fzu.edu.cn}

\address{Fujian Key Laboratory of Quantum Information and Quantum Optics, College
    of Physics and Information Engineering, Fuzhou University, Fuzhou, Fujian 350108, China}

\begin{abstract}
We propose a scheme for measuring topological properties in a two-photon-driven Kerr-nonlinear resonator (KNR) subjected to a single-photon modulation. The topological properties are revealed through the observation of the Berry curvature and hence the first Chern number, as a nonadiabatic response of the physical observable to the change rate of the control parameter of the modulated drive. The parameter manifold, constructed from the system's Hamiltonian that determines its dynamics constrained in the state space spanned by the even and odd cat states as two basis states, is adjusted so that the degeneracy crossing the manifold indicates a topological transition. The scheme, with such continuous variable states in mesoscpic systems, provides a new perspective for exploration of the geometry and the related topology with complex systems.  
\end{abstract}

\maketitle


Topology is originally a mathematical concept, which describes the properties of geometry that remains unchanged after changing shape continuously \cite{Schirber_physics2016}. Topology often possesses its physical counterpart and helps us understand more deeply the natural phenomena \cite {Klitzing_arcmp2017}. The discovery of topological properties in natural or artificial materials overthrows our traditional view of phases of matter \cite { Chen_science2012}. 
The novel phases inherent in the nontrivial topology can be characterized and classified by the topological invariant, for instance, the first Chern number \cite{Chern_AM1946}, which can be captured by the distribution of the Berry curvature and obtained by its integration over a closed surface \cite{Wen_2004}. The values of the first Chern numbers are discrete numbers and their jumps indicate nontrivial topological transitions in the system \cite{Wen_2004, Bernevig_2013}. The geometry-related quantum phenomena can be explored by a controllable quantum system \cite {Periwal_nature2022}, where the closed surface mentioned above is formed by the parameter space of the system's Hamiltonian \cite {Gritsev_pnas2012}. The external control of these parameters allows for dynamical manipulation of system's ground state associated with the specially appointed geometric manifold, where the Berry curvature can be extracted through the measurement of the physical observables and used for synthesis of the global topology \cite{Gritsev_pnas2012}. Such a topology can be characterized by the first Chern number, which keeps unchanged with perturbation to the deformation of the geometric manifold \cite{Berry_prsa1984}. 

One effective method to extract the Berry curvature and thus to obtain the first Chern number is to use a nonadiabatic response of the physical observables to the change rate of an external parameter \cite{Gritsev_pnas2012}. The physics involved in the procedure the Berry curvature acquired through such a dynamical method is that, curvature is itself a direct reflection of an effective force \cite{Gritsev_pnas2012, Schroer_prl2014, Roushan_nature2014}, like a point charge moving in a magnetic field undergoing the Lorentz force. For a driven two-level system, whose state dominated by its Hamiltonian, vividly depicted by a rotating Bloch vector ramping on the Bloch sphere, is at an arbitrary time positioned by the control of angles along both the latitude and longitude, $\theta$ and $\varphi$, respectively \cite{Berry_prsa1984}.   
Under the change of $\theta$ with the rate $\dot{\theta}=\frac{d\theta}{dt}$, the state feels an effective force: $f_{\varphi}\propto \Dot{\theta} B_{\theta \varphi}+O(\dot{\theta})$, where $B_{\theta \varphi}$ is the Berry curvature and $O(\dot{\theta})$ describes higher-order terms  \cite{Gritsev_pnas2012}. Such a force can generally be characterized by measurements of the physical observables of the system:
$\left\langle f_{\varphi} \right\rangle=-\left\langle \partial_{\varphi}H	\right\rangle$, with the system's Hamiltonian $H$ given by 
\begin{equation}\label{e1}
H/\hbar = \vec{D}\cdot \vec{\sigma},  
\end{equation} 
where $\hbar$ is the reduced Plank constant, $\vec{D}$ is the vector form of the drive field and $\vec{\sigma}$ is the Pauli matrix, whose components along the three axes are given by $\sigma_k$ ($k=x,y,z$). We look back at the referred perhaps simplest model which possesses topological properties introduced by Haldane \cite{Roushan_nature2014, Haldane_prl1988}:
\begin{equation}\label{e2}
	H_{Hal}/\hbar = (m_0-m_t)\sigma_z + \nu_F(k_x\sigma_x+k_y\sigma_y),
\end{equation}
where $m_0$ and $m_t$ are the effective mass and the second-neighboring tunnelling rate, respectively, $k_x$ and $k_y$ are the momentum-space coordinates, and $\nu_F$ is the Fermi velocity. The inherent topological behavior is that the system is in a trivial insulating phase if $m_0/m_t>1$, while for $m_0/m_t<1$ it is in a topological phase. Such a topological behavior can be reconstructed by recasting (\ref{e2}) into (\ref{e1}), mapping the momentum space of a condensed-matter model to the parameter space of a driven two-level model \cite{Roushan_nature2014}. The modulated parameters of the drive field determine $\vec{D}\equiv[\Omega sin(\theta)cos(\varphi), \Omega sin(\theta)sin(\varphi), \delta_0+\delta cos(\theta)]$
of (\ref{e1}), where $\hbar[\delta_0+\delta cos(\theta)]$ is the energy difference between the drive field and the two-level system \cite{Schroer_prl2014}. Therefore, the manifold for the parameter space takes shape: an ellipsoid with center located in the $z$ direction by $\delta_0$ and with the lengths of the semi-axes represented by $\Omega$, $\Omega$ and $\delta$, respectively, implementing the model of \cite{Klitzing_arcmp2017} where the role of $m_0/m_t$ is played by $\delta_0/\delta$. The Berry curvature needed to obtain the first Chern number can be extracted by slowly sweeping $\theta$ from $0$ along a specific parameter space trajectory to reach $\pi$. For $\delta_0=0$, this corresponds to the case the system's degeneracy point is just the center of the parameter space manifold. At this case, the motion of the system, if starting from an initial state (say, the north pole), will ramp along the $\varphi$-meridian on the Bloch sphere to the final state (the south pole), not strictly but slightly deflected from this trajectory, as compared to the case with the adiabatic ramping passage along the ground state  \cite{Gritsev_pnas2012, Schroer_prl2014, Roushan_nature2014}. Note that for $\delta_0/\delta > 1$, the motion of the system will not ramp around complete meridian on the Bloch sphere  \cite{Schroer_prl2014}. The state of the system during the ramping process is then measured in time through quantum state tomography \cite{Nielsen_2000,Yang_npjqi2021}, which gives the Berry curvature
$B_{\theta \varphi} = \frac{\left\langle \partial_{\varphi}H
		\right\rangle}{\Dot{\theta}}$
. The first Chern number is obtained by the integration of $B_{\theta \varphi}$ over the whole parameter space manifold:
$C_1= \frac{1}{2\pi}\int_0^{\pi}d\theta \int_0^{2\pi}B_{\theta\varphi}d\varphi$, where  
$C_1 = \pm 1$ ($C_1 = 0$) is for $\vert \delta_0/\delta \vert < 1$ ($\vert \delta_0/\delta \vert > 1$), which casts light upon the fixed position of the parameter space manifold, that is, encircling or not encircling the system's degeneracy \cite{Berry_prsa1984}.  

Through the method of nonadiabatic response, the measurements of the Berry curvature and the first Chern number have been conducted with a superconducting qubit, with the use of the closed manifold in the parameter space of the two-level system Hamiltonian \cite{Schroer_prl2014, Roushan_nature2014,Tan_prl2019, Wang_scpma2018}. The similar measurement with such a method has been implemented by use of two coupled superconducting qubits and revealed an interaction-induced topological phase transition \cite{Roushan_nature2014}. Recently, the related explorations through such a nonadiabatic response method has been extended to the system's temporal evolution, showing that the change of the Berry curvature provides a potential signature for tracking system's quantum states \cite{zelin_lpl2017,zelin_prd2017,zelin_epjplus2021}. All the above mentioned studies are based on the discrete two-level systems \cite{Schroer_prl2014, Roushan_nature2014,Tan_prl2019, Wang_scpma2018,zelin_lpl2017,zelin_prd2017,zelin_epjplus2021}. One question may arise whether such cases can also be applied to those of the continuous variable systems \cite{Serafini_2017,Albert_prl2016,Braunstein_rmp2005}. In this study, we address this question by exploiting the two-photon-driven KNR subjected to a single-photon modulation as a paradigm \cite{Wielinga_pra1993,Goto_pra2016,puri_npjqi2017,Grimm_nature2020}. 

For the two-photon-driven KNR, on the assumption that the frequency of the two-photon drive is twice the frequency of the resonator, and in the frame rotating with the frequency of the resonator, the system's Hamiltonian can be written as \cite{Wielinga_pra1993,Goto_pra2016,puri_npjqi2017,Grimm_nature2020}
\begin{equation}\label{e3}
\hat{H}_0/\hbar = -K \hat{a}^{\dag 2} \hat{a}^2 +P(\hat{a}^{\dag 2} +\hat{a}^2).
\end{equation}
In the above Hamiltonian of (\ref{e3}), $\hat{a}^{\dag}$ and $\hat{a}$ are the photon creation and annihilation operators, K is the strength of the Kerr nonlinearity, and P is the strength of the two-photon drive. The Hamiltonian of (\ref{e3}) possesses two degenerate eigenstates $\vert \pm \alpha\rangle$ with energy $K\alpha^4$, where $\alpha = \sqrt{P/K}$. This means that the even and odd cat states 
\begin{equation}\label{e4}
|C_{\alpha}^{\pm}\rangle=N_{\alpha}^{\pm}(|\alpha \rangle \pm |-\alpha \rangle),
\end{equation}
with the normalization coefficients $N_{\alpha}^{\pm}=1/\sqrt{2 \pm 2e^{-2|\alpha|^2}}$, are also the degenerate eigenstates of this Hamiltonian. Such states of (\ref{e4}) are referred to as the Kerr-cat qubit basis states \cite{Grimm_nature2020}. Expanded by infinite dimensional Fock states, they are belong to the continuous variable series \cite{Albert_prl2016}. Note that the action of the annihilation operator on one of these states generates the other: $\hat{a} |C_{\alpha}^{\pm}\rangle = \alpha |C_{\alpha}^{\mp}\rangle$, for certain $\alpha$ satisfying $e^{-2|\alpha|^2} << 1$ (See Appendix), meaning that one photon loss exchanges the two states rather than destroys them, in contrast to the cases with the discrete variable basis states \cite{Andersen_np2015,Wen_photonics2021,Yang_photonics2022}. Such a prominent feature makes them effective candidates for certain purposes of quantum computation \cite{Albert_prl2016,Xu_prl2020,Ofek_nature2016,Sun_nature2014}. We will show hereafter that, the application of a modulated single-photon drive to the two-photon-driven KNR can induce a nonadiabatic response of the physical observable to the change rate of the specific drive parameter, realizing the effective extraction of the Berry curvature and thus the first Chern number. 

For the two-photon-driven KNR with a pair of degenerate continuous variable cat qubit basis states as the eigenstates, the key for inducing the nonadiabatic response processes as the cases done with the discrete variable systems is to arbitrarily control the system's dynamics within the state space $\{|C_{\alpha}^{+}\rangle, |C_{\alpha}^{-}\rangle\}$ \cite{Grimm_nature2020}.  

In order to do so, we apply a single-photon modulation to the two-photon-driven KNR, whose control Hamiltonian can be written as 
\begin{equation}\label{eq4}
\hat{H}_1/\hbar = \Lambda \hat{a}^{\dag}\hat{a} + \Omega \hat{a}^{\dag} + \Omega^* \hat{a},
\end{equation}
where $\Lambda$ is the detuning between the drive frequency and the resonator photon frequency, and $\Omega$ is the drive Rabi frequency, which should satisfy $\vert \Omega \vert \ll 4K\vert \alpha \vert^2$ so as to validate the control condition mentioned above \cite{Grimm_nature2020}. Therefore, restricted within the state space $\lbrace |C_{\alpha}^{+}\rangle, |C_{\alpha}^{-}\rangle\rbrace$, the two-photon driven KNR is arbitrarily controlled through the domination of (\ref{eq4}), functioning  analogous to the case for the discrete variable two-level system:
\begin{equation}\label{e6}
\tilde{H}_1/\hbar =\Omega_x\tilde{\sigma}_x + \Omega_y\tilde{\sigma}_y + \Omega_z \tilde{\sigma}_z,
\end{equation}
where $\Omega_{x,y,z}$ are dependent on the Rabi frequency $\Omega$ and the detuning $\Lambda$. The distinction here lies in that, $\tilde{\sigma_j}$ $(j = x, y, z)$ are the new Pauli operators defined in the state space $\lbrace |C_{\alpha}^{+}\rangle, |C_{\alpha}^{-}\rangle\rbrace$, obtained through the projection of the unity operator $\mathcal P = \vert C_{\alpha}^+\rangle \langle C_{\alpha}^+ \vert +\vert C_{\alpha}^-\rangle \langle C_{\alpha}^- \vert$:  $\mathcal P (\hat{a}^{\dag} + \hat{a}) \mathcal P \sim \tilde{\sigma_x}$, $\mathcal P (\hat{a}^{\dag} - \hat{a}) \mathcal P \sim \tilde{\sigma_y}$, $\mathcal P \hat{a}^{\dag} \hat{a} \mathcal P \sim \tilde{\sigma_z}$ (see Appendix). 

\begin{figure}[htbp] 
	\centering
	\includegraphics[width=3.4in]{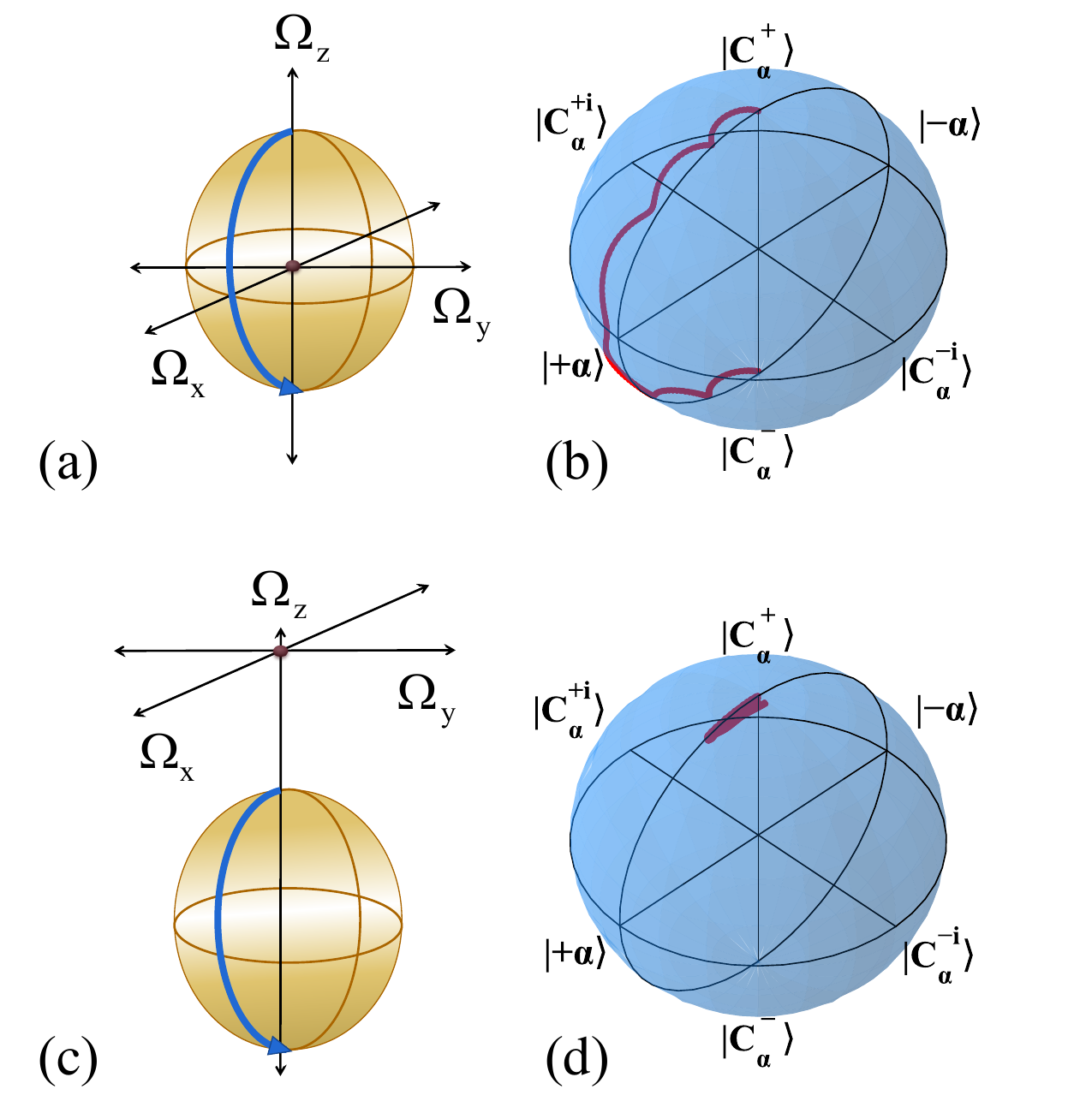}
	\caption{(Color online) The process for extracting the Berry curvature in order to get the first Chern number, exhibiting the parametric ramping trajectory (blue solid circle arc arrow) with the manifold (yellow) spanned by the Hamiltonian's parameters and correspondingly the system's state evolution route (red solid circle wave arc) moving on the Bloch sphere. (a) The degenerate point is in the middle of the manifold: $\Delta = 0$; (b) The degenerate point is outside the manifold: $\Delta = 2\Omega_{zo}$.}
	\label{fig1}
\end{figure} 

To construct a symmetric parameter manifold for the Hamiltonian of (\ref{e6}), we set the control parameters of the modulation drive as   
\begin{eqnarray}\label{4} 
\begin{aligned}
\Omega_x&=\Omega_{x_o}\sin\theta\cos\phi,\\ 
\Omega_y&=\Omega_{y_o}\sin\theta\sin\phi,\\
\Omega_z&=\Omega_{z_o}\cos\theta+\Delta.
\end{aligned}
\end{eqnarray}
All the related parameters $\Omega_{j_o}$ $(j = x,y,z)$, $\Delta$, $\theta$ and $\phi$ can be determined by the Rabi frequency $\Omega$ (including its amplitude, phase) and the detuning $\Lambda$. Notice that the Rabi frequency of the modulation drive should be predetermined by the amplitude of the two-photon drive as $\alpha$ $\infty$ $\sqrt{P}$ \cite{Grimm_nature2020}. In Fig. \ref{fig1} (a) and (c), we show the control parameter manifold, an ellipsoid with its equatorial radii along the x-axis and the y-axis corresponding to $\Omega_{x_o}$ and $\Omega_{y_o}$, respectively, while its polar radius along the z-axis corresponding to $\Omega_{z_o}$. The change of longitude-direction angle $\theta \in [0,\pi]$ and latitude-direction angle $\phi \in [0,2\pi]$ outline the whole control of the system. The manifold center is dependent on $\Delta$, which plays a key role for reflecting the topological properties related to system's control dynamics.   

The system's state is controlled to evolve towards the specific path under the setting of the parameters $\theta$ and $\phi$, through the control of the modulated single-photon drive. In Fig. \ref{fig1} (b) and (d), we show the Bloch sphere characterizing the system's state evolution, confined within the state space $\{\vert C_{\alpha}^+\rangle, \vert C_{\alpha}^-\rangle\}$. Both $\theta$ and $\phi$ are initialized to be zero. The evolution path can in principle be arbitrarily preset. Here it is chosen to follow the case that, $\theta$ is set to be a linear function of the ramp time T \cite{Schroer_prl2014,Roushan_nature2014,zelin_lpl2017}, i.e., $\theta$=$\pi $t/T, while $\phi$ is kept constantly during the whole process. The response of the continuous variable system state to the nonadiabatic manipulations of the Hamiltonian of (\ref{e6}), which keeps perturbative with respective to the Hamiltonian of (\ref{e3}), results in the so called geometric force \cite{Gritsev_pnas2012,Roushan_nature2014}: $\langle F_{\phi}\rangle=\langle \psi_0|F_{\phi}|\psi_0 \rangle-\dot{\theta}B_{\theta \phi}+\mathcal{O}(\dot{\theta})$, where $|\psi_0\rangle $ is the ground state of the system. If T is long enough, the higher-order terms $\mathcal{O}(\dot{\theta})$ are ignored \cite{Gritsev_pnas2012}. In such a case, the Berry curvature $B_{\theta \phi}$is linearly responsed and can be extracted from the geometric force $F_{\phi} =-\partial_{\phi}\tilde{H}_1$, which is given by      
\begin{equation}\label{6}
B_{\theta \phi}=\frac{\Omega_y \langle \tilde{\sigma}_{y}\rangle \sin \theta}{\dot{\theta}}.		
\end{equation}
Integrating $B_{\theta \phi}$ over the manifold spanned by the parameter space ($\theta$, $\phi$) readily gives the first Chern number
\begin{equation}\label{7}
C_{1}=\int B_{\theta \phi} d\theta,
\end{equation}
which is expressed in a $\theta$-dependent form for the sake of computational convenience, as the whole integral is cylindrically symmetric about the z axis \cite{Roushan_nature2014}. As the system has a degenerate point at $\tilde{H}_1 = 0$,  the integration of $B_{\theta \phi}$ over the whole parameter manifold only gives $C_1$ of either 1 (-1) or 0 \cite{Wen_2004, Bernevig_2013}, dependent on the relative position between the degenerate point and the manifold \cite{Schroer_prl2014}. When the degenerate point is inside the manifold $(|\Delta|<|\Omega_{z_o}|)$, as shown in Fig. \ref{fig1} (a), whose case corresponds to  $\Delta$ = 0, it gives $C_1$ = 1 (starting with $\vert C_{\alpha}^+\rangle$) or $C_1 = -1$ (starting with $\vert C_{\alpha}^-\rangle$); while the degenerate point is outsides the manifold $(|\Delta|>|\Omega_{z_o}|)$,  as seen from Fig. \ref{fig1} (c), which takes the instance of $\Delta_{z_o}$ = 2$|\Omega_{z_o}|$, it obtains $C_1 = 0$. The two cases $|\Delta|<|\Omega_{z_o}|$ and $|\Delta|>|\Omega_{z_o}|$ correspondingly reflect the dynamics process of the continuous variable system's state, which wraps or does not wrap the Bloch sphere \cite{Schroer_prl2014}, as depicted in Fig. \ref{fig1} (b) and (d), respectively.  

\begin{figure}[h!] 
	\centering
	\includegraphics[width=3.4in]{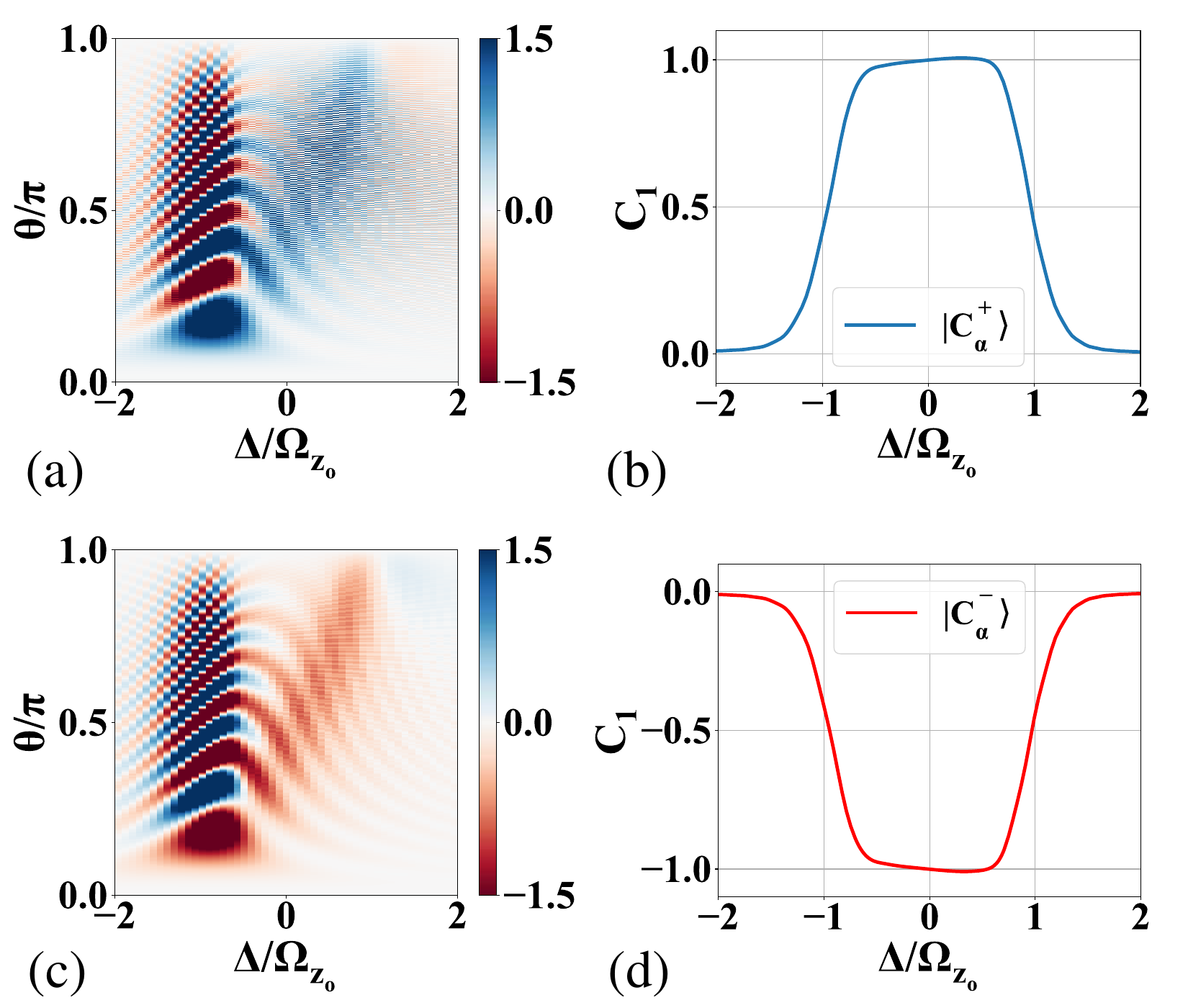}
	\caption{(Color online) The extracted Berry curvature $B_{\theta \phi}$ and the corresponding first Chern number $C_1$ for the nonadiabatic response processes. (a), (c) The Berry curvature $B_{\theta \phi}$ versus $\theta \in [0,\pi]$ and $\Delta \in [-2,2]\Omega_{z_o}$, for the system's initial state $\vert C_{\alpha}^+\rangle$ and $\vert C_{\alpha}^-\rangle$, respectively. (b), (d) The first Chern number $C_1$ versus $\Delta \in [-2,2]\Omega_{z_o}$, corresponding to (a), (c), respectively. The related parameters are chosen as follows: $K = 2\pi \times 6.7$ MHz, $P = 2\pi \times 4.3$ MHz, $\Omega_{x_o} = \Omega_{y_o}= 2\pi \times 0.25$ MHz, $\Omega_{z_o} = 2\pi\times0.34$ MHz, and $T= 10$ $\mu$s.  }
	\label{fig2}
\end{figure}

To quantitatively characterize the dynamics process of the nonadiabatic response, in Fig. \ref{fig2}, we show the extracted $B_{\theta \phi}$ (versus $\theta$ and $\Delta$) and the first Chern number $C_1$ (versus $\Delta$), by considering two cases for the driven KNR's initial state, $\vert C_{\alpha}^+\rangle$ for Fig. \ref{fig2} (a) and (b), $\vert C_{\alpha}^-\rangle$ for Fig. \ref{fig2} (c) and (d), by choosing a set of parameters:  $K= 2\pi \times 6.7$ MHz, $P= 2\pi \times 4.3$ MHz, $\Omega_{x_o} = \Omega_{y_o}=2\pi \times 0.25$ MHz, and $\Omega_{z_o}=2\pi\times0.34$ MHz. The comparison between Fig. \ref{fig2} (a) and (c) clearly shows that, the Berry curvature extracted from the nonadiabatic response of $\tilde{\sigma}_y$ to $\dot{\theta}$, exhibits complementarity about positive and negative values. This indicates that, the system starting from the two different initial states $\vert C_{\alpha}^+\rangle$ and $\vert C_{\alpha}^-\rangle$, respectively, ramps along the ground state $\vert \psi_0\rangle$ and its `mirror ground state' during the dynamic response process. This pair of the ground states are the dressed eigenstates of the combined Hamiltonians of (\ref{e3}) and (\ref{e6}): manipulated by the Hamiltonian (\ref{e6}) and perturbative over the Hamiltonian (\ref{e3})'s constrained state space $\{\vert C_{\alpha}^+\rangle, \vert C_{\alpha}^-\rangle\}$. The oscillation of the Berry curvatures also hints the transformation of the system's state back and forth between $\vert C_{\alpha}^+\rangle$ and $\vert C_{\alpha}^-\rangle$, in the opposite direction within the two complementary ground states. The amplitudes of such oscillations reach the peak values when the degenerate point happens to cross the manifold for $\Delta/\Omega_{z_o} = -1$, though this is not the case for $\Delta/\Omega_{z_o} = 1$. With the increase of $\Delta/\Omega_{z_o}$, the degenerate point gradually enters into the manifold, the oscillations of the Berry curvature trend smoothly, become positive-value-dominated and negative-value-dominated for the cases with initial $\vert C_{\alpha}^+\rangle$ and $\vert C_{\alpha}^-\rangle$, respectively, and begin to reverse when the degenerate point happens to cross the manifold at the symmetric end: $\Delta/\Omega_{z_o} = 1$. 

\begin{figure}[htbp] 
	\centering
	\includegraphics[width=3.4in]{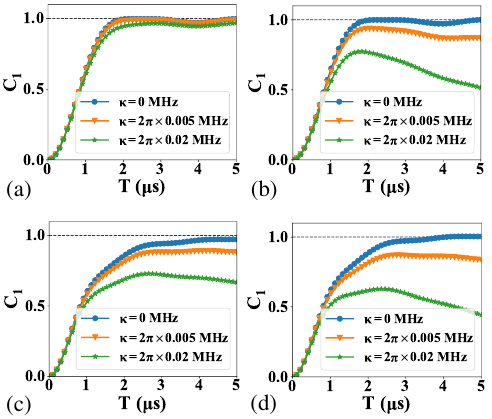}
	\caption{(Color online) The first Chern number $C_1$ under different duration T for the nonadiabatic response processes, for $\Delta = 0$, $K = 2\pi \times 6.7$ MHz, $\kappa = 2\pi \times (0, 0.005, 0.02)$ MHz, $n_{th}=0$, and $\kappa_d=0$. (a), (b) $P = 2\pi \times 4.3$ MHz, $\Omega_{x_o} = \Omega_{y_o}= 2\pi \times 0.25$ MHz, and $\Omega_{z_o} = 2\pi\times0.34$ MHz, obtained from the cases with the system's initial state $\vert C_{\alpha}^+\rangle$ and $\vert C_{\alpha}^-\rangle$, respectively. (c), (d) $P = 2\pi \times 9.6$ MHz, $\Omega_{x_o} = \Omega_{y_o}= 2\pi \times 0.35$ MHz, and $\Omega_{z_o} = 2\pi\times0.65$ MHz, corresponding to the initial $\vert C_{\alpha}^+\rangle$ and $\vert C_{\alpha}^-\rangle$, respectively.}
	\label{fig3}
\end{figure} 

In Fig. \ref{fig2} (b) and (d), we show the first Chern number $C_1$ for different $\Delta$, integrating the corresponding Berry curvature shown in Fig. \ref{fig2} (a) and (c) over the whole parameter manifold. It obviously shows that, the first Chern number $C_1$ only takes 1 (-1) and 0, corresponding to the dynamics process of the system ramping within the ground (`mirror ground') state, and jumps between 1 (-1) and 0 when the system's degenerate point crosses the parameter manifold, indicating a topological transition in the system. The stability of the first Chern number holding for a specific range of $\Delta$ makes clear that it is robust against the certain perturbation to the system. Notice that this robustness is also independent of deformations of the closed manifold that does not cross the degeneracy \cite{Schroer_prl2014}. The induced not so sharp topological transitions, as shown in Fig. \ref{fig2} (b) and (d), as compared to the ideal ones \cite{Wen_2004, Bernevig_2013}, is mainly due to the not long enough duration for the linear response processes with the chosen parameters \cite{Schroer_prl2014,Roushan_nature2014,zelin_lpl2017}.  

\begin{figure}[htbp] 
	\centering
	\includegraphics[width=3.4in]{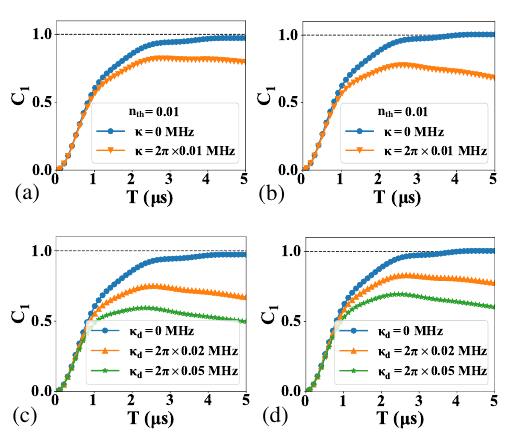}
	\caption{(Color online) The first Chern number $C_1$ under different duration T for the nonadiabatic response, with the preset parameters: $\Delta = 0$, $K = 2\pi \times 6.7$ MHz, $P = 2\pi \times 9.6$ MHz, $\Omega_{x_o} = \Omega_{y_o}= 2\pi \times 0.35$ MHz, and $\Omega_{z_o} = 2\pi\times0.65$ MHz, for the cases corresponding to the system's initial state: (a) (c) $\vert C_{\alpha}^+\rangle$; (b) (d) $\vert C_{\alpha}^-\rangle$. The other parameters for (a) (b) are: $\kappa = 2\pi \times (0, 0.01)$ MHz, $n_{th}=0.01$, $\kappa_d=0$; while those for (c) (d) are: $\kappa_d = 2\pi \times (0, 0.02, 0.05)$ MHz, $\kappa = 0$.}
	\label{fig4}
\end{figure}

In order to check the influences of the decoherence of the system coupling to the environment, the numerical simulations are conducted with the Lindblad master equation \cite{Gardiner_2000,Grimm_nature2020} 
\begin{eqnarray}
\frac{d}{dt} \hat{\rho}(t) &=& -i/\hbar [\hat{H}_0+\hat{H_1}, \hat{\rho}(t)] + \kappa (1+n_{th}) D[\hat{a}]\rho(t) \nonumber \\
 && + \kappa D[\hat{a}^{\dag}]\rho(t) + \kappa_d D[\hat{a}^{\dag} \hat{a}]\rho(t), 
\end{eqnarray}
where the Lindblad super-operator is defined for the quantum operator $\hat{o}$ ($\hat{o} \in \hat{a}, \hat{a}^{\dag}, \hat{a}^{\dag}\hat{a}$) as $D[\hat{o}]\hat{\rho} = \hat{o}\hat{\rho}(t)\hat{o}^{\dag} -\frac{1}{2}\hat{o}^{\dag}\hat{o}\rho(t) -\frac{1}{2}\rho(t)\hat{o}^{\dag}\hat{o}$, with $\kappa$ and $\kappa_d$ being the rates of the single-photon loss and the dephasing of the KNR, respectively, and $n_{th}$ denoting its equilibrium thermal photon occupation number \cite{Grimm_nature2020}. In Fig. \ref{fig3}, we show the first Chern number $C_1$ integrated over the whole parameter manifold located at $\Delta = 0$, as a function of the duration $T$ for different nonadiabatic response processes, for the initial $\vert C_{\alpha}^+\rangle$ and $\vert C_{\alpha}^-\rangle$, respectively, with $\kappa = 2\pi \times (0, 0.005, 0.02)$ MHz, $n_{th} =0$, and $\kappa_d = 0$. In the calculations, the related parameters for Fig. \ref{fig3} (a) and (b) are chosen to be: $K = 2\pi \times 6.7$ MHz, $P = 2\pi \times 4.3$ MHz, $\Omega_{x_o} = \Omega_{y_o}= 2\pi \times 0.25$ MHz, and $\Omega_{z_o} = 2\pi\times0.34$ MHz; while those for Fig. \ref{fig3} (c) and (d) are chosen as: $K = 2\pi \times 6.7$ MHz, $P = 2\pi \times 9.6$ MHz, $\Omega_{x_o} = \Omega_{y_o}= 2\pi \times 0.35$ MHz, and $\Omega_{z_o} = 2\pi\times0.65$ MHz. It clearly shows that the influences of the system's single-photon loss on the nonadiabatic response processes during which the system's ramping along the two ground state manifolds are asymmetric, due to the unequal-weight-exchange between the two cat qubit basis states induced by the single-photon annihilation during the ramping processes. Such asymmetric influences ease with increasing the amplitude of the coherent state component involved in the ground states: from $\alpha \sim 0.8$ in Fig. \ref{fig3} (a) and (b) to $\alpha \sim 1.2$ in Fig. \ref{fig3} (c) and (d), and vanish for $\alpha$ satisfying $e^{-2|\alpha|^2} \ll 1$.
In Fig. \ref{fig4} (a) and (b), we show the influences of the thermal photon occupation on the first Chern number $C_1$ under different duration T for the system's ramping during the nonadiabatic response processes, with the similar parameters chosen as those for Fig. \ref{fig3} (c) and (d). It is obvious that the thermal photon occupation aggravates the weakening of the Chern number, as it not only accelerates the photon dissipation escaping from the ground manifold within the state space $\{\vert C_{\alpha}^+\rangle, \vert C_{\alpha}^-\rangle\}$, but also mixes up the intrinsic state components of the ground state due to the photon creation induced by the thermal photon occupation. We further show the influences of the KNR's dephasing on the first Chern number $C_1$ in Fig. \ref{fig4} (c) and (d), which indicate that the dephasing affects the system's nonadiabatic responses starting from the different state $\vert C_{\alpha}^+\rangle$ and $\vert C_{\alpha}^-\rangle$ asymmetrically. Such asymmetric weakening on the Chern number for $\vert C_{\alpha}^+\rangle$ and $\vert C_{\alpha}^-\rangle$, however, is contrary to the effects induced by the single-photon loss, as shown in Fig. \ref{fig3}. The dephasing affects the Chern number with initial state $\vert C_{\alpha}^-\rangle$ more seriously than that with $\vert C_{\alpha}^+\rangle$, and becomes balanced for larger $\alpha$. 

With the recently developed or sooner to be improved techniques with superconducting circuits \cite{Grimm_nature2020}. such kinds of topological transitions seem to be observable once the nonadiabatic response with the long enough duration is allowed and the related parameters required are met. Several requirements are needed for the extraction of the Berry curvature so as to obtain first Chern number. (a) The first is the initialization of the system in $\vert C_{\alpha}^+\rangle$ or $\vert C_{\alpha}^-\rangle$, achieved by preparing the driven KNR in the vacuum (single-photon) state and then by mapping such a Fock state to the even (odd) cat state, through ramping on the two-photon drive \cite{Grimm_nature2020}; (b) The second is the control of the nonadiabatic response of the physical observables to the change rate of the control parameters, accomplished by controlling the single-photon modulation on the driven KNR. (c) The third is the state tomography of the driven KNR's state during the nonadiabatic response process, finished by mapping the even (odd) cat state back to the Fock state via ramping off the two-photon drive at each measurement time, then by applying tomography drives to the Fock qubit \cite{Grimm_nature2020}, and finally followed by dispersive readout \cite{Wallraff_nature2004}.         

In summary, we have proposed a scheme for observation of topological transitions in the driven KNR subjected to a single-photon modulation. The combined Hamiltonian of the system is tailored to construct an ellipsoid parameter manifold, while the system's dynamics ramps within the confined state space where the even and odd cat states act as the continuous variable basis states. The Berry curvature is extracted from the nonadiabatic response of the physical observable to the change rate of the modulation drive's control parameter, and integrated over the whole parameter manifold to obtain the first Chern number. The topological transitions characterizing the jumps of the Chern number are revealed through moving the manifold to cross the system's degeneracy. Recent development of techniques with superconducting transmon under parametric driving seems to provide a good physical platform for observation of the kind of topological transitions \cite{Grimm_nature2020}.       

\medskip
\textbf{Supporting Information} \par 
For the cat-qubit state subspace $\{\vert C_{\alpha}^+\rangle, \vert C_{\alpha}^-\rangle\}$, where $\vert C_{\alpha}^{\pm}\rangle = N_{\alpha}^{\pm} (\vert \alpha\rangle + \vert -\alpha\rangle)$, with $N_{\alpha}^{\pm} = (2 \pm 2e^{-2|\alpha|^2})^{-1/2}$ being the normalization coefficients, the projector for the cat-qubit state subspace can be defined as 
\begin{equation}
	\mathcal P = \vert C_{\alpha}^+\rangle \langle C_{\alpha}^{+}\vert + \vert C_{\alpha}^-\rangle \langle C_{\alpha}^-\vert. 	
\end{equation}

Through the relations:
\begin{eqnarray}
	\hat{a} \vert C_{\alpha}^+\rangle = \alpha  N_{\alpha}^{+} (\vert \alpha\rangle - \vert -\alpha\rangle) \equiv \alpha \frac{N_{\alpha}^{+}}{N_{\alpha}^{-}} \vert C_{\alpha}^-\rangle, \\
	\hat{a} \vert C_{\alpha}^-\rangle = \alpha  N_{\alpha}^{-} (\vert \alpha\rangle + \vert -\alpha\rangle) \equiv \alpha \frac{N_{\alpha}^{-}}{N_{\alpha}^{+}} \vert C_{\alpha}^+\rangle, \\
	\langle C_{\alpha}^+\vert \hat{a}^{\dag} = (\langle \alpha \vert - \langle -\alpha \vert)N^{+}_{\alpha}\alpha^* \equiv \alpha^* \frac{N^{+}_{\alpha}}{N^{-}_{\alpha}}\langle C_{\alpha}^{-}\vert, \\
	\langle C_{\alpha}^-\vert \hat{a}^{\dag} = (\langle \alpha \vert + \langle -\alpha \vert)N^{-}_{\alpha}\alpha^* \equiv \alpha^* \frac{N^{-}_{\alpha}}{N^{+}_{\alpha}}\langle C_{\alpha}^{+}\vert,
\end{eqnarray} 
the new Pauli operators defined on the pair of continuous variable cat states subspace can be obtained by the projection of the new unity operator $\mathcal P$ onto the operators $\hat{a}-\hat{a}^{\dag}$, $\hat{a}+\hat{a}^{\dag}$, and $\hat{a}^{\dag}\hat{a}$, respectively, shown as below:
\begin{eqnarray}
	&&\mathcal P (\hat{a} + \hat{a}^{\dag}) \mathcal P \nonumber \\ 
	&=& (\vert C_{\alpha}^{+}\rangle \langle C_{\alpha}^+\vert + \vert C_{\alpha}^{-}\rangle \langle C_{\alpha}^-\vert) (\hat{a} + \hat{a}^{\dag}) (\vert C_{\alpha}^{+}\rangle \langle C_{\alpha}^+\vert + \vert C_{\alpha}^{-}\rangle \langle C_{\alpha}^-\vert)  \nonumber \\ 
	&=&4\alpha N^+ N^- (\vert C_{\alpha}^{+}\rangle \langle C_{\alpha}^-\vert + \vert C_{\alpha}^{-}\rangle \langle C_{\alpha}^+\vert)  \nonumber \\
	&=& 4\alpha N^+ N^- \tilde{\sigma}_x,
\end{eqnarray}

\begin{eqnarray}
	&&\mathcal P (\hat{a} - \hat{a}^{\dag}) \mathcal P \nonumber \\
	&=& (\vert C_{\alpha}^{+}\rangle \langle C_{\alpha}^+\vert + \vert C_{\alpha}^{-}\rangle \langle C_{\alpha}^-\vert) (\hat{a}^{\dag} - \hat{a}) (\vert C_{\alpha}^{+}\rangle \langle C_{\alpha}^+\vert + \vert C_{\alpha}^{-}\rangle \langle C_{\alpha}^-\vert)   \nonumber \\ 
	&=& -4i\alpha N^+ N^- e^{-2|\alpha|^2}(-i\vert C_{\alpha}^{+}\rangle \langle C_{\alpha}^-\vert +i \vert C_{\alpha}^{-}\rangle \langle C_{\alpha}^+\vert) \nonumber \\
	&=& -4i\alpha N^+ N^- e^{-2|\alpha|^2}\tilde{\sigma}_y, 
\end{eqnarray}

\begin{eqnarray}
	&&\mathcal P \hat{a}^{\dag}\hat{a} \mathcal P \nonumber \\
	&=& (\vert C_{\alpha}^+\rangle \langle C_{\alpha}^{+}\vert + \vert C_{\alpha}^-\rangle \langle C_{\alpha}^-\vert) \hat{a}^{\dag}\hat{a} (\vert C_{\alpha}^+\rangle \langle C_{\alpha}^{+}\vert + \vert C_{\alpha}^-\rangle \langle C_{\alpha}^-\vert) \nonumber \\
	&=&|\alpha|^2 \frac{{N_{\alpha}^+}^2}{{N_{\alpha}^-}^2} \vert C_{\alpha}^+\rangle \langle C_{\alpha}^{+}\vert + |\alpha|^2 \frac{{N_{\alpha}^-}^2}{{N_{\alpha}^+}^2} \vert C_{\alpha}^-\rangle \langle C_{\alpha}^-\vert   \nonumber \\
	&=&|\alpha|^2 \frac{{N_{\alpha}^+}^2}{{N_{\alpha}^-}^2}\frac{1}{2}(\tilde{I}+\tilde{\sigma_z}) + |\alpha|^2 \frac{{N_{\alpha}^-}^2}{{N_{\alpha}^+}^2}\frac{1}{2}(\tilde{I}-\tilde{\sigma_z}) \nonumber \\
	&=&|\alpha|^2{N_{\alpha}^+}^2{N_{\alpha}^-}^2(4 + 4e^{-4|\alpha|^2})\tilde{I} + 8|\alpha|^2{N_{\alpha}^+}^2{N_{\alpha}^-}^2e^{-2|\alpha|^2}\tilde{\sigma}_z,  \nonumber \\
\end{eqnarray}
where $\tilde{I}=\mathcal P$.

\begin{acknowledgments}
This work was supported by the National Natural Science Foundation of China under Grand Nos. 12274080, 11875108, and 11874114.
\end{acknowledgments}

\bibliography{template}

\end{document}